# Observational constraints on the neutron star mass distribution

Lee Samuel Finn
Department of Physics and Astronomy,
Northwestern University, Evanston, Illinois 60208-3112
(September 21, 1994)

Radio observations of neutron star binary pulsar systems have constrained strongly the masses of eight neutron stars. Assuming neutron star masses are uniformly distributed between lower and upper bounds $m_l$ and $m_u$, the observations determine with 95% confidence that $1.01 < m_l/M_\odot < 1.34$ and $1.43 < m_u/M_\odot < 1.64$. These limits give observational support to neutron star formation scenarios that suggest that masses should fall predominantly in the range $1.3 < m/M_\odot < 1.6$, and will also be important in the interpretation of binary inspiral observations by the Laser Interferometer Gravitational-wave Observatory.

PACS numbers: 97.60.Jd, 04.40.Dg, 97.10.Nf

*a. Introduction and Motivation* Radio observations of four neutron star - neutron star (ns-ns) binary pulsar systems have constrained the masses of eight neutron stars, and all the masses lie close to $1.35\,M_\odot$. This coincidence suggests that natural formation mechanisms restrict the range of ns masses more than limitations due to the nuclear and super-nuclear equation of state. This suggestion also arises in theoretical studies of core collapse supernovae [1]. Here I explore the statistical significance of this coincidence, with the goal of determining nature's limits on ns masses: in particular, modeling the ns mass distribution as uniform between upper and lower bounds $m_u$ and $m_l$, I use the observations to determine the joint probability distribution of $m_u$ and $m_l$.

While noted before [2], the coincidence in measured ns masses has never been subjected to a statistical analysis that treats the observations jointly and accounts for the statistical uncertainties in the mass determinations. Here I use Bayesian statistical techniques to combine the separate observations and determine the probability distribution of the mass bounds.

A second motivation for this study is the anticipated observation of binary ns inspiral by the Laser Interferometer Gravitational-wave Observatory (LIGO) [3]. LIGO will be sensitive to ns masses in inspiraling binaries [4,5], and interpretation of its observations will require an accurate assessment of our knowledge of the mass distribution in ns-ns binaries [6]. This work is meant to begin that assessment.

*b. Summary of relevant observations* Observations of binary pulsars 1913+16 and 1534+12 have determined for each system the total mass $M$ and the pulsar companion mass $m_c$ [7,8]. In both cases the companion is believed to be a ns. Similarly, observations of binary pulsars 2127+11C and 2303+46 have determined for each $M$ and the mass function $f$ [9,2]. Assuming a probability density $P(\cos i) = 1/2$ for the orbital inclination angle $i$ limits the mass of the pulsar and its companion, which (for these two systems) is believed to be a ns. Here I assume that the measured $f$, $M$, and $m_c$ are distributed independently and normally about the actual $\widehat{f}$, $\widehat{M}$, and $\widehat{m}_c$, where the distribution variances are given by the reported $1\sigma$ uncertainties in $f$, $M$, and $m_c$:

$$P(x|\widehat{x},\mathcal{I}) = \exp\left[-\frac{1}{2}\left(\frac{x-\widehat{x}}{\sigma_x}\right)^2\right]/\sqrt{2\pi}\sigma_x \qquad (1)$$

where $x$ is one of $f$, $M$, or $m_c$, and $\mathcal{I}$ represents other, unenumerated assumptions that characterize the observations. Tables I and II give the values adopted here for $f$, $m_c$, $M$, $\sigma_f$, $\sigma_{m_c}$ and $\sigma_M$ of these four systems.

TABLE I. The values adopted for the total mass $M$, the companion mass $m_c$, and the standard error of each ($\sigma_M$ and $\sigma_{m_c}$) of PSR1913+16 and PSR1534+12 [7,8].

| System  | $M/M_\odot$ | $\sigma_M/M_\odot$ | $m_c/M_\odot$ | $\sigma_{m_c}/M_\odot$ |
|---------|-------------|--------------------|---------------|------------------------|
| 1913+16 | 2.82827     | $4 \times 10^{-5}$ | 1.442         | 0.003                  |
| 1534+12 | 2.679       | 0.003              | 1.36          | 0.03                   |

TABLE II. The values adopted for the mass function $f$, the total mass $M$, and the standard error of each ($\sigma_f$ and $\sigma_M$) of PSRs 2127+11C and 2303+46 [9,2,17].

| System      | $f/M_\odot$ | $\sigma_f/M_\odot$ | $M/M_\odot$ | $\sigma_M/M_\odot$   |
|-------------|-------------|--------------------|-------------|----------------------|
| PSR2127+11C | 0.15285     | $1.8 \times 10^{-4}$ | 2.706     | $3.6 \times 10^{-3}$ |
| PSR2303+46  | 0.246287    | $6.7 \times 10^{-6}$ | 2.57      | 0.08                 |

The relationship between $f$, $M$, $m_c$, $\widehat{f}$, $\widehat{M}$, and $\widehat{m}_c$ is not as simple as equation (1); however, for small variances the Gaussian approximation is a good one, and in the absence of a detailed description of the fitting procedure that determines $f$, $m_c$, and $M$ from the observations the assumption of statistical independence is reasonable. Additionally, numerical investigations show that the final 95% confidence intervals for $m_l$ and $m_u$ are largely independent of the choice of $P(m_c, M|\widehat{m}_c, \widehat{M}, \mathcal{I})$ and $P(f, M|\widehat{f}, \widehat{M}, \mathcal{I})$.

No other observed binary pulsar is known to have a ns companion. Both $f$ and $M$ have also been determined for PSRs 1855+09 and 1802-07 [10,2]; however, in these systems the companion is thought to be a white dwarf. The constraints these observations place on the ns mass depend on the uncertain distribution of white dwarf masses in pulsar-white dwarf binaries; consequently, I do not consider these systems in making my estimates.

In addition to ns-ns binary pulsars, where the masses are determined through observations of the pulsar Doppler velocity curve with precision sufficient to observe relativistic effects, ns masses have also been determined in several X-ray binary systems [11,12]. These masses are estimated from Keplerian-order observations of *both* the pulsar and the companion Doppler velocity curves, and an estimate of the orbital inclination angle based on the eclipse duration. The derived masses are more uncertain than for any of the ns-ns binaries considered here; additionally, X-ray binary neutron stars may constitute a sub-population with a different mass distribution than ns-ns binaries. In order to maintain a homogeneous sample of neutron stars as well as to avoid the complications inherent in the modeling of the inclination angle I have excluded the X-ray binary observations from the sample considered here (however, see sec. d).

*c. Probability and statistics of neutron star masses*
Could we examine all neutron stars, we would know the exact distribution of their masses. Lacking complete knowledge, we can examine an incomplete sample and a family of hypothetical distributions and calculate the conditional probability that the sample is drawn from a member of the family. As the sample size decreases or the observations become less precise, we are less able to discriminate between distributions that differ only slightly. For small or imprecise samples, only the grossest features of the distribution can be characterized. From observations of four ns-ns binaries all we can realistically hope to learn of the ns mass distribution is its mean and extent. Here I consider a simple family of distributions that captures these gross features. As more observations become available, our understanding can be refined by refining the family of distributions.

Thus, suppose that ns masses are uniformly distributed between upper and lower bounds $m_u$ and $m_l$, so that $P(m|m_l, m_u, \mathcal{I}) = 1/(m_u - m_l)$ when $m_l < m < m_u$, and 0 otherwise. In this Letter I find $P(m_l, m_u|\{g\}, \mathcal{I})$, where $\{g\}$ represents the ns-ns binary observations from which the masses are determined.

Using Bayes Law of conditional probabilities, $P(m_l, m_u|\{g\}, \mathcal{I})$ may be factorized:

$$P(m_l, m_u|\{g\}, \mathcal{I}) = \frac{P(\{g\}|m_l, m_u, \mathcal{I})}{P(\{g\}|\mathcal{I})} P(m_l, m_u|\mathcal{I}). \quad (2)$$

The probability density $P(m_l, m_u|\mathcal{I})$ represents our prior knowledge of the upper and lower bounds $m_u$ and $m_l$. Impose here only the most conservative theoretical constraints: causality and general relativity together provide that neutron stars cannot be more massive than $M_u \simeq 3\,\mathrm{M}_\odot$ [13,14], and our understanding of the subnuclear density equation of state provides that they are not less massive than $M_l \simeq 0.1\,\mathrm{M}_\odot$ [15]. As a result, $M_u > m_u > m_l > M_l$. Theory providing no further clear guidance, assume that within these constraints all pairs $(m_l, m_u)$ are equally likely; thus

$$P(m_l, m_u|\mathcal{I}) = 2/(M_u - M_l)^2 \quad (3)$$

for $M_l < m_l < m_u < M_u$ and 0 otherwise.

For a given set of observations $\{g\}$, $P(\{g\}|\mathcal{I})$ is a constant whose value must be such that

$$1 = \int_{M_l}^{M_u} dm_l \int_{m_l}^{M_u} dm_u \, P(m_l, m_u|\{g\}, \mathcal{I}). \quad (4)$$

Thus, once $P(\{g\}|m_l, m_u, \mathcal{I})$ is known, $P(m_l, m_u|\{g\}, \mathcal{I})$ is determined through this normalization integral and there is no need to find $P(\{g\}|\mathcal{I})$ separately.

To evaluate $P(\{g\}|m_l, m_u, \mathcal{I})$, note that the observations of each binary pulsar system are independent; consequently,

$$P(\{g\}|m_l, m_u, \mathcal{I}) = \prod_n P(g_n|m_l, m_u, \mathcal{I}) \quad (5)$$

where $g_n$ represents the observation of system $n$. The form of $P(g_n|m_l, m_u, \mathcal{I})$ is determined by the character of the observation $g_n$. For PSRs 1913+16 and 1534+12, we need the joint probability distribution $P(m_c, M|m_l, m_u, \mathcal{I})$ of the observed companion mass $m_c$ and total mass $M$ for fixed $m_l$ and $m_u$. Using Bayes law, this distribution can be expressed as an integral over $P(m_c, M|\widehat{m_c}, \widehat{M}, \mathcal{I})$ (cf. eq. 1):

$$P(m_c, M|m_l, m_u, \mathcal{I}) =$$
$$\iint d\widehat{m_c}\, d\widehat{M}\; P(m_c, M|\widehat{m_c}, \widehat{M}, \mathcal{I}) P(\widehat{m_c}, \widehat{M}|m_l, m_u, \mathcal{I})$$
$$(6)$$

where $P(\widehat{m_c}, \widehat{M}|m_l, m_u, \mathcal{I}) = (m_u - m_l)^{-2}$.

For PSRs 2127+11C and 2303+46, we need the distribution $P(f, M|m_l, m_u, \mathcal{I})$ of the observed mass function $f$ and total mass $M$ for fixed $m_l$ and $m_u$. This distribution can be expressed as an integral over $P(f, M|\widehat{f}, \widehat{M}, \mathcal{I})$ (cf. eq. 1):

$$P(f, M|m_l, m_u, \mathcal{I}) =$$
$$\iint d\widehat{f}\, d\widehat{M}\; P(f, M|\widehat{f}, \widehat{M}, \mathcal{I}) P(\widehat{f}, \widehat{M}|m_l, m_u, \mathcal{I}) \quad (7)$$

To evaluate $P(\widehat{f}, \widehat{M}|m_l, m_u, \mathcal{I})$, write it as

$$P(\widehat{f}, \widehat{M}|m_l, m_u, \mathcal{I}) =$$
$$P(\widehat{f}|\widehat{M}, m_l, m_u, \mathcal{I}) P(\widehat{M}|m_l, m_u, \mathcal{I}). \quad (8)$$

The two probability densities on the right-hand side can be calculated separately:

$$P(\widehat{M}|m_l, m_u, \mathcal{I}) = \frac{\max\left[0, \min\left(\widehat{M} - m_l, m_u\right) - \max\left(\widehat{M} - m_u, m_l\right)\right]}{(m_u - m_l)^2} \quad (9a)$$

$$P(\widehat{f}|\widehat{M}, m_l, m_u, \mathcal{I}) = \frac{1}{3} \frac{\left(\sin^{-1} x_1 - \sin^{-1} x_0\right) \left(\widehat{M}/\widehat{f}\right)^{2/3}}{\min\left(\widehat{M} - m_l, m_u\right) - \max\left(\widehat{M} - m_u, m_l\right)} \quad (9b)$$

where

$$x_0^2 = \max\left[0, 1 - \frac{\left(\widehat{f}\widehat{M}^2\right)^{2/3}}{\min\left[m_u, \max\left(\widehat{M} - m_u, m_l\right)\right]^2}\right] \quad (9c)$$

$$x_1^2 = \max\left[0, 1 - \frac{\left(\widehat{f}\widehat{M}^2\right)^{2/3}}{\max\left[m_l, \min\left(\widehat{M} - m_l, m_u\right)\right]^2}\right]. \quad (9d)$$

In the event $M$ is unknown but we believe the system consists of two neutron stars, we can write

$$P(f|m_l, m_u, \mathcal{I}) = \int dM \, P(f, M|m_l, m_u, \mathcal{I}) \quad (10)$$

Either of expressions (8) or (10) may be useful in interpreting observations of binary pulsars systems in the context of a model for the pulsar mass distribution.

*d. Discussion* Figure 1 shows contours of constant $P(m_l, m_u|\{g\}, \mathcal{I})$ containing 95% (solid line) and 68% (dotted line) of the total probability. Figures 2 and 3 show the same contours if the survey is restricted to those systems where the individual masses are known (PSRs 1913+16 and 1534+12) or to those where only the mass function and the total mass are known (PSRs 2127+11C and 2303+46).

Summarizing figure 1, the 95% confidence intervals for the lower and upper bounds $m_l$ and $m_u$ when considered separately, are $1.01 < m_l/M_\odot < 1.34$ and $1.43 < m_u/M_\odot < 1.64$, and the maximum likelihood values of $m_l$ and $m_u$ are $1.29 \, M_\odot$ and $1.45 \, M_\odot$. The 95% confidence intervals for the mass range $\Delta = m_u - m_l$ and the total mass $M$, when considered separately, are $0.10 < \Delta/M_\odot < 0.51$ and $2.46 < M/M_\odot < 2.88$, and the maximum likelihood values of $\Delta$ and $M$ are $0.16 \, M_\odot$ and $2.73 \, M_\odot$.

Quite independently of the observations, numerical calculations of gravitational collapse supernovae have led Woosley and Weaver to "predict" ns masses in the range 1.15–2.0 $M_\odot$, with a preponderance between 1.3 and 1.6 $M_\odot$ [1]. This prediction is certainly consistent with the range derived here.

While I have excluded X-ray binaries from the analysis presented here (cf. sec. b), it is worth noting that the inferred mass of the Vela X-1 ns is greater than $1.50 \, M_\odot$ with 95% probability [11]. An upper limit $m_u$ greater than this is consistent with our results even at the 68% confidence level; however, $m_u$ greater than the most likely value of the Vela X-1 ns mass estimated by [11] ($1.85\,M_\odot$) lies outside the 95% confidence interval estimated here.

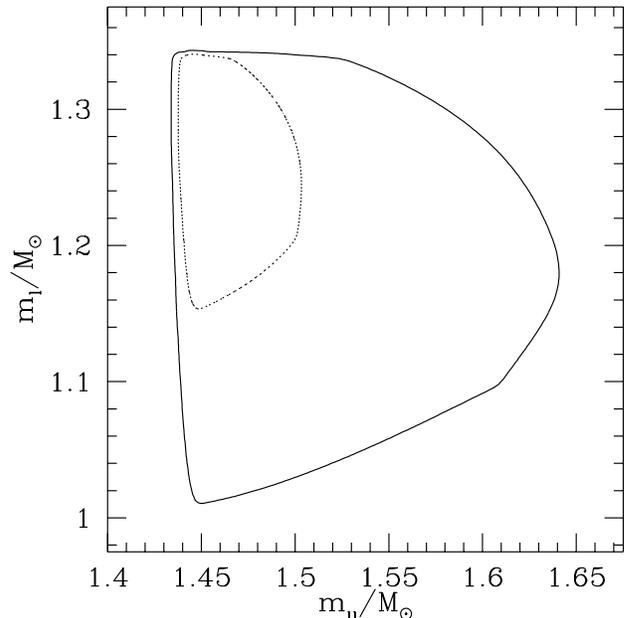

FIG. 1. Assuming ns masses are uniformly distributed between $m_l$ and $m_u$, observations of PSRs 1534+12, 1913+16, 2127+11C, and 2303+46 determine the joint probability distribution for $m_l$ and $m_u$. Shown here are contours enclosing regions of 68% (dotted) and 95% (solid) of this distribution.

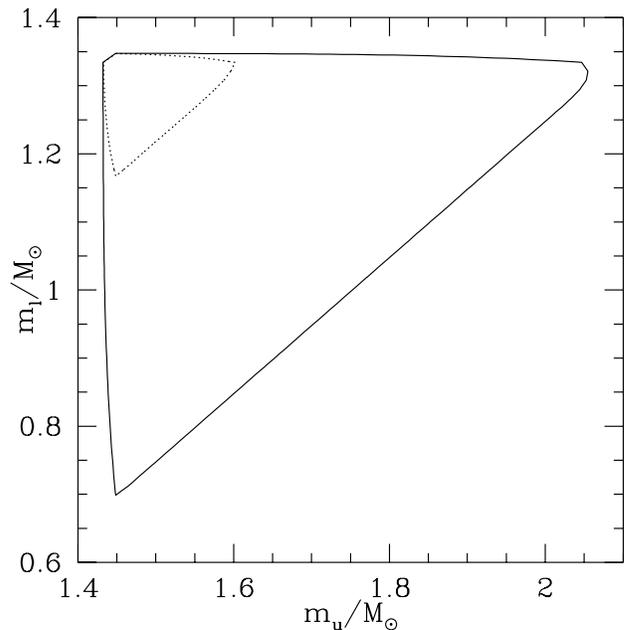

FIG. 2. As in figure 1, except that the contours are based on the constraints provided by observations of PSRs 1534+12 and 1913+16.

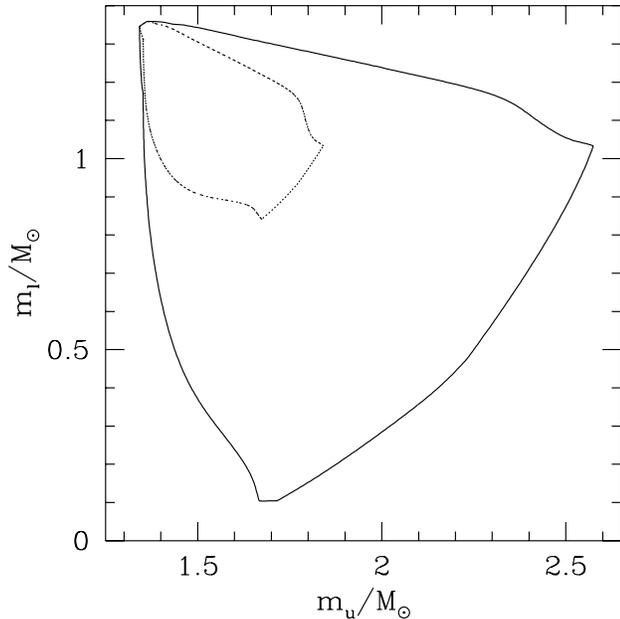

FIG. 3. As in figure 1, except that the contours are based on the constraints provided by observations of PSRs 2127+11C and 2303+46.

LIGO will be extremely sensitive to the "chirp mass" $[\mathcal{M} = (m_1 m_2)^{3/5}(m_1 + m_2)^{-1/5}]$ of an inspiraling binary system, but less sensitive to the individual masses [4,5]. In the same manner that our prior knowledge that ns masses could in no case be greater than $3\,M_\odot$ or less than $0.1\,M_\odot$ played a role in this analysis (cf. eq. 2, 3 and intervening text), so our understanding of the range and distribution of ns masses gained from this exercise will play a role in determining the confidence intervals for measurements of $\mathcal{M}$, $m_1$, and $m_2$ using LIGO observations. An accurate assessment of our prior knowledge is especially important in determining when a signal is sufficiently strong that it refines our understanding as opposed to affirming our existing prejudices.

The ns sample used in this survey is highly selected: only components of ns-ns binary pulsar systems are included. In addition to the exclusion of X-ray binaries and isolated neutron stars, this means that half the sample are pulsars. It is believed that isolated, non-millisecond pulsars are a fraction $10^{-4}$ of neutron stars, and that ns-ns binary pulsars number approximately 1/10 of these [16]. Since the mass distribution of the ns sub-population considered here may not be representative of neutron stars generally, application of these results to isolated neutron stars must be made cautiously. Without good estimates of the how the mass distributions of these different populations may differ it is not possible to estimate the effects of these selection biases. Nevertheless, it is clear that the homogeneous subset of neutron stars considered here has, with high confidence, a range of masses restricted in a way that our understanding of ns and binary system formation and evolution do not, but eventually must, confront.

I am grateful to Sterl Phinney and Steve Thorsett for helpful conversations. This work was supported by the Alfred P. Sloan Foundation and the National Science Foundation (PHY 9308728).

[1] S. E. Woosley and T. A. Weaver, in *The Structure and Evolution of Neutron Stars*, edited by D. Pines, R. Tamagaki, and S. Tsuruta (Addison-Wesley, Redwood City, California, 1992), pp. 235–249.
[2] S. E. Thorsett, Z. Arzoumanian, M. M. McKinnon, and J. H. Taylor, Astrophys. J. Lett. **405**, L29 (1993).
[3] A. Abramovici *et al.*, Science **256**, 325 (1992).
[4] L. S. Finn and D. F. Chernoff, Phys. Rev. D **47**, 2198 (1993).
[5] C. Cutler and É. Flanagan, Phys. Rev. D **49**, 2658 (1994).
[6] L. S. Finn, Observing binary inspiral with LIGO, To appear in proceedings of the Lanczos International Centenary, 1994, Northwestern University preprint NU-GR-7.
[7] J. H. Taylor and J. M. Weisberg, Astrophys. J. **345**, 434 (1989).
[8] A. Wolszczan, Nature (London) **350**, 688 (1991).
[9] T. A. Prince, S. B. Anderson, S. R. Kulkarni, and A. Wolszczan, Astrophys. J. Lett. **374**, L41 (1991).
[10] M. F. Ryba and J. H. Taylor, Astrophys. J. **371**, 733 (1991).
[11] P. C. Joss and S. A. Rappaport, Annu. Rev. Astron. Astrophys. **22**, 537 (1984).
[12] F. Verbunt, Annu. Rev. Astron. Astrophys. **31**, 93 (1993).
[13] C. E. Rhoades and R. Ruffini, Phys. Rev. Lett. **32**, 324 (1974).
[14] J. B. Hartle, Phys. Rep. **46**, 201 (1978).
[15] S. L. Shapiro and S. A. Teukolsky, *Black Holes, White Dwarfs, and Neutron Stars* (Wiley, New York, 1983).
[16] R. Narayan, T. Piran, and A. Shemi, Astrophys. J. Lett. **379**, L17 (1991).
[17] S. E. Thorsett, private communication, 1994.